\documentclass[sigconf]{acmart}
\AtBeginDocument{%
  }
\setcopyright{acmlicensed}
\copyrightyear{2024}
\acmYear{2024}
\setcopyright{rightsretained}
\acmConference[VRST '24]{30th ACM Symposium on Virtual Reality Software and Technology}{October 9--11, 2024}{Trier, Germany}
\acmBooktitle{30th ACM Symposium on Virtual Reality Software and Technology (VRST '24), October 9--11, 2024, Trier, Germany}\acmDOI{10.1145/3641825.3689700}
\acmISBN{979-8-4007-0535-9/24/10}

\begin{document}

\title{Single Vs Dual: Influence of the Number of Displays on User Experience within Virtually Embodied Conversational Systems}


\author{Navid Ashrafi}
\affiliation{
  \institution{Immersive Reality Lab, \hbox{Hochschule Hamm-Lippstadt}}
  \city{Hamm}
  \country{Germany}
}

\author{Francesco Vona}
\affiliation{%
  \institution{Immersive Reality Lab, \hbox{Hochschule Hamm-Lippstadt}}
  \city{Hamm}
  \country{Germany}
}

\author{Philipp Graf}
\affiliation{%
  \institution{University of Applied Sciences Munich}
  \city{Munich}
  \country{Germany}
}

 \author{Philipp Harnisch}
\affiliation{%
  \institution{Technical University of Berlin}
  \city{Berlin}
  \country{Germany}
}

\author{Sina Hinzmann}
\affiliation{%
  \institution{Immersive Reality Lab, \hbox{Hochschule Hamm-Lippstadt}}
  \city{Hamm}
  \country{Germany}
}

\author{\hbox{Jan-Niklas Voigt-Antons}}
\affiliation{%
  \institution{Immersive Reality Lab, \hbox{Hochschule Hamm-Lippstadt}}
  \city{Hamm}
  \country{Germany}
}
\email{Jan-Niklas.Voigt-Antons@hshl.de}

\renewcommand{\shortauthors}{Ashrafi et al.}

\begin{abstract}
The current research evaluates user experience and preference when interacting with a patient-reported outcome measure (PROM) healthcare application displayed on a single tablet in comparison to interaction with the same application distributed across two tablets. 
We conducted a within-subject user study with 43 participants who engaged with and rated the usability of our system and participated in a post-experiment interview to collect subjective data. 
Our findings showed significantly higher usability and higher pragmatic quality ratings for the single tablet condition. 
However, some users attribute a higher level of presence to the avatar and prefer it to be placed on a second tablet. 

\end{abstract}


\keywords{Virtual agents, tablet, PROMs, healthcare applications}
  


\maketitle

\begin{figure}[ht]
\centerline{\includegraphics[width=.6\columnwidth, height=.25\textheight]{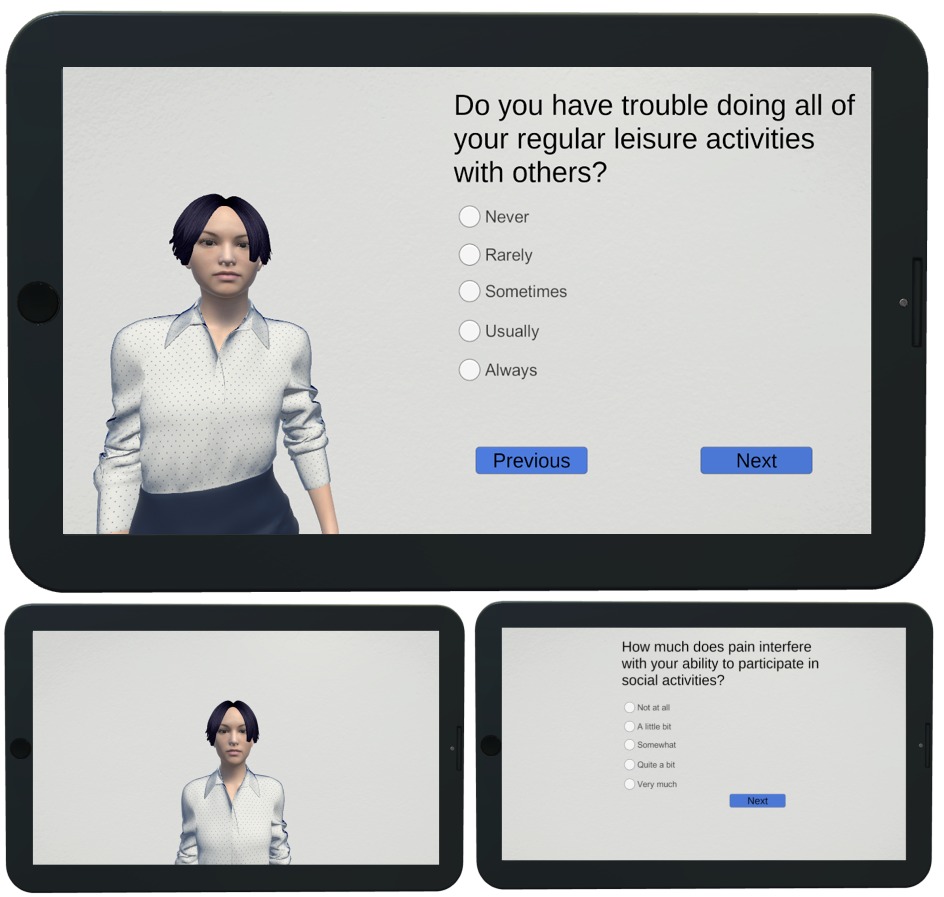}}
\centering

\caption{Two experiment conditions: (top) \textit{Single display} where a virtual avatar is integrated with the questionnaire, and (bottom) \textit{dual display} where the virtual avatar is separated from the questionnaire and the devices are synchronized using an MQTT server providing the same functionality as in the \textit{single display} condition.}
\label{setup}
\end{figure}
\section{Introduction}

Completion of long, paper-based Patient-reported Outcomes Measures (PROMs) in healthcare can introduce a great burden\cite{long2022} and lead patients to skip items. This will pose a significant concern for the Quality of User Experiences (QoE) \cite{kohn}. Virtual characters in healthcare applications can potentially enhance the QoE by improving the level of assistance offered to patients and leveraging the sense of social presence\cite{Lee2004}. 
The perceived trustworthiness of an avatar is shaped by both warmth and competence, hence, avatars that exude warmth and competence are generally favored, as these qualities instill a greater sense of credibility in the avatars \cite{10}. 
In the context of healthcare applications requiring data disclosure, an open question is the placement of the virtual assistant on a separate. The use of a virtual avatar on a separate tablet involves proxemic aspects concerning the spatial relations between entities \cite{Hall1966}. Disproportionate proxemic distance can provoke strong emotional reactions, making proxemic rules a sensitive, yet effective design tool. 
In our case, proxemic relations are crucial since displaying the avatar on a separate tablet changes its position in the space. 
Cross-device and multi-device systems have been explored in various contexts, including spatially-aware techniques for document sharing \cite{redle2} and multiple tablets for data visualization \cite{wig}. However, existing research has primarily addressed scenarios with more than two devices and collaborative use cases rather than exploring one versus two devices. Notably, no research has been conducted addressing virtual avatars in multi-device settings. 

We designed an exploratory study to assess the QoE when collecting a 26-item health-related quality of life questionnaire (WHOQOL-BREF) \cite{whool} in two (\textbf{single} and \textbf{dual}) display conditions featuring a virtual avatar (Figure \ref{setup}). 
To the best of our knowledge, this study is the first to address the proxemic aspect of multi-device, avatar-based healthcare applications. 
This study was conducted as a usability evaluation test for a multimodal assistance system (incorporating a virtual assistant displayed on a separate tablet) developed to be tested in rehabilitation centers with patients suffering from mental illnesses to assist with the collection of PROM data.

\section{Methodology}

We used the \textit{Unity engine} to develop two versions of an application differing only in the display of the virtual avatar. In the single tablet condition, a female virtual avatar integrated into the same tablet as the questionnaire items was displayed in the opening scene. Upon tapping "Start", the first question appeared, with the avatar reading it to the user. Users could tap their answers, navigate to the next question, or go back to the previous one (Figure \ref{setup}). In the dual tablet condition, the virtual avatar appeared on another tablet. The tablets were synchronized via a local MQTT server, providing the same functionality as in the single display condition. The virtual avatar (female avatar with a warm and competent voice and appearance) was designed using \textit{Daz Studio} \cite{daz2022}. Synthetic voices were generated using Google's text-to-speech cloud service and synchronized with the avatar through the \textit{SALSA Lipsynch} \cite{salsa2022}. Participants (43 in total, 22 males and 21 females) interacted with both conditions equally in a randomized order.
We collected demographic data 
and the Affinity for Technology Interaction (ATI) \cite{ati} information. To assess the QoE we used the standard System Usability Scale (SUS) \cite{sus} and User Experience Questionnaire (UEQ) \cite{ueq} scales after each condition. Participants also took part in a brief post-experiment interview where they were asked to provide more detailed feedback on the usability and trust aspects.

\section{Results \& Discussion} 
The results from the usability assessments reveal a clear preference trend: 36 of 43 participants preferred the condition of a single table in terms of general usability and pragmatic quality. In contrast, four participants chose the dual tablet setting, and three individuals were undecided. The significantly better scores for the single-tablet condition may be because most people are accustomed to using their smartphones (single devices) daily. We interpret this finding to be compatible with the existing literature on multi-device interaction, which shows a preference for fewer devices \cite{plank}. We used Analysis of Variance (ANOVA) to assess the influence of demographic variables and technological affinity on the QoE. Participants with academic backgrounds rated the one tablet condition significantly higher compared to non-academic participants ($F(3, 35) = 3.67, p = .021$), particularly in terms of general usability of SUS, and UEQ hedonic (how interesting, exciting, inventive, and usual) factors. Notably, the order of presentation in the first condition did reveal a significant effect ($F(3, 35) = 3.95, p = .016$), influencing both SUS ($F = 5.03, p = .031, \eta^2_p = .0013$) and UEQ\_prag scores ($F = 5.93, p = .018, \eta^2_p = .0015$). 
Additionally, female participants consistently rated both conditions (SUS: $F(1, 41) = 4.74, p = .035, \eta^2_p = .32$; UEQ: $F(1, 41) = 5.11, p = .029, \eta^2_p = .33$) higher than their male counterparts across all scores. However, when gender was considered as a covariate, it did not demonstrate significance. 

The change of focus caused by the use of two tablets (mentioned by users who preferred a single tablet) is a clear usability limit. Our evaluations affirm the importance of proxemic rules: The situation must always be set up and arranged in such a way that a handy interaction can take place. The avatar in this case must therefore be arranged at an appropriate proxemic distance to the questionnaire. While we consider digital social agents to be confined to the physical space provided by hardware, such as a tablet, there might be an opportunity that in the future, these systems will become more ubiquitous. Therefore, digital social agents can also accompany users without the restriction of physical hardware \cite{Luria2019}. 

\bibliographystyle{ACM-Reference-Format}
\bibliography{main}

\end{document}